\documentclass[%
reprint,
superscriptaddress,
pra,
amsmath,amssymb,
aps,
]{revtex4-2}

\usepackage{bm}
\usepackage[caption=false]{subfig}
\usepackage{braket}
\usepackage{amsfonts}
\usepackage{graphicx,color}
\usepackage{booktabs,multirow}

\newcommand{\YMdel}[1]{\textcolor{red}}

\begin{document}
\title{
Quantum Thermodynamics applied for Quantum Refrigerators cooling down a qubit
}
\author{Hideaki Okane}
\affiliation{Research Center for Emerging Computing Technologies (RCECT), 
National Institute of Advanced Industrial Science and Technology (AIST), 
1-1-1, Umezono, Tsukuba, Ibaraki 305-8568, Japan}

\author{Shunsuke Kamimura}
\affiliation{Research Center for Emerging Computing Technologies (RCECT), 
National Institute of Advanced Industrial Science and Technology (AIST), 
1-1-1, Umezono, Tsukuba, Ibaraki 305-8568, Japan}
\affiliation{Faculty of Pure and Applied Sciences, University of Tsukuba, 
Tsukuba 305-8571, Japan}
\author{Shingo Kukita}
\affiliation{Department of Physics, Kindai University, 
Higashi-Osaka 577-8502, Japan and}
\author{Yasushi Kondo}
\affiliation{Department of Physics, Kindai University, 
Higashi-Osaka 577-8502, Japan and}
\author{Yuichiro Matsuzaki}
\email{matsuzaki.yuichiro@aist.go.jp}
\affiliation{Research Center for Emerging Computing Technologies (RCECT), 
National Institute of Advanced Industrial Science and Technology (AIST), 
1-1-1, Umezono, Tsukuba, Ibaraki 305-8568, Japan}
\affiliation{NEC-AIST Quantum Technology Cooperative Research Laboratory,
National Institute of Advanced Industrial Science and Technology (AIST), 
Tsukuba, Ibaraki 305-8568, Japan}
\date{\today}

\begin{abstract}
We discuss a quantum refrigerator to increase the ground state 
probability of a target qubit whose energy difference between the ground and excited 
states is less than the thermal energy
of the environment. We consider two types of quantum refrigerators: 
(1) one extra qubit with frequent pulse operations and
(2) two extra qubits without them. 
These two types of refrigerators are evaluated 
from the viewpoint of quantum thermodynamics. 
More specifically,
we calculate the heat removed from the target qubit, the work done for the system, and the coefficient of performance (COP), the ratio between the heat ant the work.
We show that the COP of the second type outperforms that of the first type.
Our results are useful to design a high-performance quantum refrigerator cooling down a qubit. 
\end{abstract}

\maketitle

\section{Introduction}
\label{sec:intro}
Thermodynamics has traditionally explained macroscopic behavior of classical systems.
For example, a refrigerator is essential not only for daily life but also for academic purposes.
We need a refrigerator 
for realizing interesting phenomena, such as superfluidity, superconductivity,
Bose-Einstein Condensation, and so on \cite{tilley2019superfluidity}.

On the other hand, many efforts have been devoted to extending the conventional thermodynamics to the quantum one \cite{doi:10.1080/00107514.2016.1201896, binder2018thermodynamics, 10.1088/2053-2571/ab21c6, PRXQuantum.3.020101}. Such an extension is called quantum thermodynamics. 
Quantum thermodynamics define the thermodynamic properties of microscopic systems: 
heat and work.
Quantum thermodynamics helps to understand a quantum refrigerator, which increases the population of the ground state of a target quantum system.
To quantify the performance of a quantum refrigerator, 
quantities such as coefficient of performance (COP) and cooling power are evaluated~\cite{PhysRevA.76.032325, Weimer_2008, PhysRevLett.114.100404, e18020048,  https://doi.org/10.48550/arxiv.2109.14056}.

Electron spin resonance (ESR) is an important technique to 
detect target electron spins, which gives information of various materials
\cite{lund2011principles}. 
ESR has recently been performed with several types of quantum detectors such as a superconducting circuit 
\cite{PhysRevB.86.064514,doi:10.1063/1.4940978,Bienfait2016,PhysRevLett.118.037701,
doi:10.1063/1.5002540,PhysRevX.7.041011,PhysRevMaterials.2.011403,Toida2019, 
doi:10.1063/5.0004322,doi:10.1063/1.4921231,PhysRevB.97.024419,
doi:10.1063/1.5006693,doi:10.1063/1.5144722} 
and nitrogen vacancy (NV) centers \cite{Grinolds2013, deLange2012, 
PhysRevB.87.195414, PhysRevLett.113.197601, doi:10.1063/1.4963717, 
doi:10.1126/sciadv.1701116, doi:10.1063/5.0006014}. 
Such a kind of detector can also be utilized as a quantum refrigerator to polarize target spins~\cite{liu2014protection,london2013detecting}. 
To improve the sensitivity of the ESR, polarizing the target spins is essential.
Therefore, evaluation of a quantum refrigerator should contribute to a further improvement of the ESR sensitivity.

In this work, we evaluate the performance of two types of quantum refrigerators: (1) one extra qubit with frequent pulse operations and (2) two extra qubits without them. The former corresponds to the realized scheme~\cite{liu2014protection,london2013detecting}. The latter is a newly proposed method in this paper, in which one of the extra qubits is for quantum control and the other is for the heat release to the environment.
We calculate the heat removed from the target qubit, the work done for the system, and the COP for these two approaches.
According to the analysis of them based on quantum thermodynamics, 
we find that the latter outperforms the former in terms of the COP.

The rest of this paper is organized as follows. 
\S~\ref{sec:def} reviews the definition of heat 
and  work in the bipartite quantum system, based on Ref.~\cite{Alipour2016}.
We present two models with and without frequent 
reset in \S~\ref{sec:model} and analyze their performance 
from the viewpoint of quantum thermodynamics in \S~\ref{sec:result}. 
\S~\ref{sec:summary} summarizes our results.
In appendix \ref{app:a}, we explain the possible experimental realization of our scheme with current technology.
In appendix \ref{app:b}, we explain the detail of the conventional protocol.
In appencix \ref{app:c}, we calculate the work necessary for a qubit initialization.
We take the natural unit system and thus we omit $\hbar$ and $k_{\mathrm B}$ in this paper.

\section{The definition of work and heat between interacting bipartite systems}
\label{sec:def}
Let us review the definition of work and heat transferred between an interacting bipartite quantum system (a target and extra qubits) based on quantum thermodynamics
\cite{Alipour2016}.
In order to quantify the cooling effect, we need to evaluate a heat transfer from the target qubit to the extra qubit. 

To calculate heat and work in a quantum system, we need 
to define an internal energy.
In a single system with a Hamiltonian $H$ and a density matrix $\rho$, 
the internal energy is defined as $\mathrm{Tr}(\rho H)$.
By differentiating the internal energy with time, we obtain 
$\frac{d}{dt}(\mathrm{Tr}(\rho H))=\mathrm{Tr}(\dot{\rho} H)+\mathrm{Tr}(\rho \dot{H})$.
Here, $\mathrm{Tr}(\dot{\rho} H)$ and $\mathrm{Tr}(\rho \dot{H})$ 
are defined as the heat and the work, respectively.
However, when we consider a bipartite system, 
it is not straightforward to define the internal energy for each system.
We consider a bipartite system consisting of system A and B.
The Hamiltonian is given as
\begin{equation}
	H=H_\mathrm{A}\otimes I_B + I_A \otimes H_\mathrm{B}+H_\mathrm{AB},
\end{equation}
where $H_\mathrm{A}$ ($H_\mathrm{B}$) is the Hamiltonian for system A (B), 
$H_{\mathrm{AB}}$ is the interaction Hamiltonian, $I_A$ ($I_B$) denotes identity operator for system A (B). Let $\rho$ denote the density matrix of the total system.
The reduced density matrix of system A (B) is defined as $\rho_{\mathrm{A}}=\mathrm{Tr}_{\mathrm{B}}(\rho)$ ($\rho_{\mathrm{B}}
=\mathrm{Tr}_{\mathrm{A}}(\rho)$).
We introduce the correlation between system A and B as
\begin{equation}
	\chi=\rho - \rho_\mathrm{A}\otimes\rho_\mathrm{B}.
\end{equation}

The naive definition of the internal energy for system $A$ 
may be given as $\mathrm{Tr}_A\left(\rho_\mathrm{A} H_\mathrm{A}\right)$.
However, the reduced density matrix $\rho_\mathrm{A}$ 
can be accessible to the interaction Hamiltonian, namely, 
$\mathrm{Tr}_\mathrm{A}\left(\left(\rho_\mathrm{A}\otimes I_B\right) H_{\mathrm{AB}}\right)\neq 0$.
So, we should take into account the contribution from the interaction Hamiltonian 
to define the internal energy for each system.
We reconstruct the Hamiltonian 
$H=H_\mathrm{A}^{(\mathrm{eff})}\otimes I_B + I_A \otimes H_\mathrm{B}^{(\mathrm{eff})}
+H_{\mathrm{AB}}^{(\mathrm{eff})}$ 
for the reduced density matrix to be inaccessible to the effective 
interaction Hamiltonian, satisfying the following conditions
\begin{align}
\mathrm{Tr}_\mathrm{A}\left(\left(\rho_\mathrm{A}\otimes I_B\right) H_{\mathrm{AB}}^{(\mathrm{eff})}\right)\label{eff_ham_a}
&=0,\\
\mathrm{Tr}_\mathrm{B}\left(\left(I_A \otimes \rho_\mathrm{B}\right) H_{\mathrm{AB}}^{(\mathrm{eff})}\right)
&=0.\label{eff_ham_b}
\end{align}
We adopt such effective Hamiltonians $H_{\mathrm{A}}^{(\mathrm{eff})}$ and 
$H_{\mathrm{B}}^{(\mathrm{eff})}$ to define the internal energy 
for system A and B, respectively.
Thus, the heat and the work to system A is given as 
$\mathrm{Tr}_A\left(\dot{\rho}_\mathrm{A} H_{\mathrm{A}}^{(\mathrm{eff})}\right)$ 
and $\mathrm{Tr}_A\left(\rho_\mathrm{A} \dot{H}_{\mathrm{A}}^{(\mathrm{eff})}\right)$, 
respectively.
In the following, we specifically derive the effective Hamiltonian.

We show how to construct the Hamiltonian satisfying Eqs.~\eqref{eff_ham_a} and \eqref{eff_ham_b}.
The time evolution of the total density matrix is written as,
\begin{align}
	\frac{d\rho(t)}{dt}&=-i \left[H(t),\rho(t)\right],\label{bipa_master}\\
	H(t)&=H_\mathrm{A}(t)\otimes I_B + I_A \otimes H_\mathrm{B}+H_\mathrm{AB}.
\end{align}
By taking the partial trace of system B in Eq.~\eqref{bipa_master}, 
we obtain the time-evolution equation for system A,
\begin{align}
\frac{d\rho_\mathrm{A}(t)}{dt}
&=-i \left[H_\mathrm{A}'(t),\rho_\mathrm{A}(t)\right]
-i\mathrm{Tr}_B\left(\left[H_{\mathrm{AB}},\chi(t)\right]\right),\label{time_evo_A}\\
H_\mathrm{A}'(t)
&=H_\mathrm{A}(t)+\mathrm{Tr}_\mathrm{B}\left(\left(I_A \otimes \rho_\mathrm{B}(t)\right)H_{\mathrm{AB}}\right).
\end{align}
Similarly, the time-evolution equation for the system B is given as,
\begin{align}
\frac{d\rho_\mathrm{B}(t)}{dt}
&=-i \left[H_\mathrm{B}'(t),\rho_\mathrm{B}(t)\right]
-i\mathrm{Tr}_A\left(\left[H_{\mathrm{AB}},\chi(t)\right]\right),\label{time_evo_B}\\
H_\mathrm{B}'(t)
&=H_\mathrm{B}+\mathrm{Tr}_\mathrm{A}\left(\left(\rho_\mathrm{A}(t)\otimes I_B\right) H_{\mathrm{AB}}\right).
\end{align}
By using the new Hamiltonian $H_\mathrm{A}'(t)$ and $H_\mathrm{B}'(t)$, 
we can rewrite the Hamiltonian as follows,
\begin{align}
H&=H_\mathrm{A}'(t)\otimes I_B + I_A \otimes H_\mathrm{B}'(t)+H_{\mathrm{AB}}'(t),\\
H_{\mathrm{AB}}'(t)&=H_{\mathrm{AB}}
-\mathrm{Tr}_\mathrm{B}\left(\left(I_A\otimes\rho_\mathrm{B}(t)\right)H_{\mathrm{AB}}\right)\otimes I_B\\
&\qquad\qquad- I_A \otimes \mathrm{Tr}_\mathrm{A}\left(\left(\rho_\mathrm{A}(t)\otimes I_B\right)H_{\mathrm{AB}}\right),
\end{align}
where $H_{\mathrm{AB}}'(t)$ denotes the new interaction Hamiltonian.
Now, let us check whether the reduced density matrix $\rho_\mathrm{A}$ 
is inaccessible to the new interaction Hamiltonian $H_{\mathrm{AB}}'(t)$,
\begin{align}
\mathrm{Tr}_\mathrm{A}\left(\left(\rho_\mathrm{A}(t) \otimes I_B\right) H_{\mathrm{AB}}'(t)\right)
=-\mathrm{Tr}\left(\left(\rho_\mathrm{A}(t)\otimes\rho_\mathrm{B}(t)\right) H_{\mathrm{AB}}\right) I_B.
\end{align}
The reduced density matrix is still accessible to the interaction Hamiltonian.
However, since 
$-\mathrm{Tr}\left(\left(\rho_\mathrm{A}(t)\otimes\rho_\mathrm{B}(t)\right) H_{\mathrm{AB}}\right)$ 
is a scalar quantity, we can define the effective interaction Hamiltonian 
to extract the scalar part as follows,
\begin{align}
H_{\mathrm{AB}}^{(\mathrm{eff})}(t)
&=H_{\mathrm{AB}}'(t)
+\mathrm{Tr}\left(\left(\rho_\mathrm{A}(t)\otimes\rho_\mathrm{B}(t)\right) H_{\mathrm{AB}}\right)(I_A\otimes I_B).
\end{align}
This satisfies the inaccessible condition of Eqs.~\eqref{eff_ham_a} and \eqref{eff_ham_b}.
By using the effective interaction Hamiltonian, 
the total one can be rewritten as,
\begin{align}
H(t)&=H_{\mathrm{A}}^{(\mathrm{eff})}(t)\otimes I_B
+I_A \otimes H_{\mathrm{B}}^{(\mathrm{eff})}(t)+H_{\mathrm{AB}}^{(\mathrm{eff})}(t),\\
H_{\mathrm{A}}^{(\mathrm{eff})}(t)
&=H_{\mathrm{A}}'(t)-(1-\alpha)\mathrm{Tr}
\left(\left(\rho_\mathrm{A}(t)\otimes\rho_\mathrm{B}(t)\right) H_{\mathrm{AB}}\right)I_A,\\
H_{\mathrm{B}}^{(\mathrm{eff})}(t)
&=H_{\mathrm{B}}'(t)-\alpha\mathrm{Tr}
\left(\left(\rho_\mathrm{A}(t)\otimes\rho_\mathrm{B}(t)\right) H_{\mathrm{AB}}\right)I_B,
\end{align}
where an arbitrary parameter $\alpha \in \mathbb{R}$ 
is introduced for the general expression of the effective Hamiltonian.

By using the effective Hamiltonian, we define the internal energy as follows,
\begin{align}
U&=\mathrm{Tr}\left(\rho(t)H(t)\right)=U_\mathrm{A}+U_\mathrm{B}+U_\mathrm{\chi},\\
U_\mathrm{A}&=\mathrm{Tr}_A\left(\rho_\mathrm{A}(t)H_\mathrm{A}^{(\mathrm{eff})}(t)\right),\\
U_\mathrm{B}&=\mathrm{Tr}_B\left(\rho_\mathrm{B}(t)H_\mathrm{B}^{(\mathrm{eff})}(t)\right),\\
U_\mathrm{\chi}&=\mathrm{Tr}\left(\chi(t)H_\mathrm{AB}^{(\mathrm{eff})}(t)\right),
\end{align}
where $U_\mathrm{A}$ ($U_\mathrm{B}$) is the internal energy of 
the reduced density matrix $\rho_\mathrm{A}(t)$ ($\rho_\mathrm{B}(t)$), 
and $U_\mathrm{\chi}$ is the internal energy of the correlation $\chi(t)$.
The heat flux $\dot{Q}_\mathrm{A}$ ($\dot{Q}_\mathrm{B}$) 
to system A (B) is defined as,
\begin{align}
\dot{Q}_\mathrm{A}(t) 
&= \mathrm{Tr}_A\left(\dot{\rho}_\mathrm{A}(t)H_\mathrm{A}^{(\mathrm{eff})}(t)\right)\nonumber\\
&=-i\mathrm{Tr}\left(\left[H_{\mathrm{AB}},\chi(t)\right]\left(H_\mathrm{A}'(t)\otimes I_B\right)\right),
\label{def_heat}\\
\dot{Q}_\mathrm{B}(t) 
&= \mathrm{Tr}_B\left(\dot{\rho}_\mathrm{B}(t)H_\mathrm{B}^{(\mathrm{eff})}(t)\right)\nonumber\\
&=-i\mathrm{Tr}\left(\left[H_{\mathrm{AB}},\chi(t)\right]\left(I_A \otimes H_\mathrm{B}'(t)\right)\right),
\end{align}
where we used the time-evolution equation of Eqs.~\eqref{time_evo_A} and \eqref{time_evo_B}.
Note that the positive value of the heat denotes the inflow to the system.
Similarly, the heat flux to the correlation $\chi$ is also defined as,
\begin{align}
\dot{Q}_\mathrm{\chi}
&=\mathrm{Tr}\left(\dot{\chi}(t)H_\mathrm{AB}^{(\mathrm{eff})}(t)\right)
=-\left(\dot{Q}_\mathrm{A}+\dot{Q}_\mathrm{B}\right).
\end{align}
It is worth mentioning that these heat fluxes do not depend on $\alpha$.
In the heat transfer between system A and  B, 
a part of the heat is absorbed by the correlation.
Specifically, when the heat flux $-\dot{Q}_\mathrm{B}$ is released 
from system B, system A obtains the heat flux of 
$-\dot{Q}_\mathrm{B} -\dot{Q}_\mathrm{\chi}$ 
where the correlation absorbs the heat flux of $\dot{Q}_\mathrm{\chi}$.

The work rate to system A (B) is defined as,
\begin{align}
\dot{W}_\mathrm{A} 
&= \mathrm{Tr}_A\left(\rho_\mathrm{A} \dot{H}_\mathrm{A}^{(\mathrm{eff})}\right)
=\mathrm{Tr}_A\left(\rho_\mathrm{A}(t)\dot{H}_\mathrm{A}(t)\right)
\nonumber \\
&-(1-\alpha)\mathrm{Tr}\left(\left(\dot{\rho}_\mathrm{A}(t)\otimes \rho_\mathrm{B}(t)\right)H_{\mathrm{AB}}\right)\nonumber\\
&+\alpha \mathrm{Tr}\left(\left(\rho_\mathrm{A}(t) \otimes \dot{\rho}_\mathrm{B}(t)\right)H_{\mathrm{AB}}\right),\\
\dot{W}_\mathrm{B} 
&= \mathrm{Tr}_B\left(\rho_\mathrm{B} \dot{H}_\mathrm{B}^{(\mathrm{eff})}\right)
\nonumber \\
&=(1-\alpha)\mathrm{Tr}\left(\left(\dot{\rho}_\mathrm{A}(t)\otimes
\rho_\mathrm{B}(t)\right)H_{\mathrm{AB}}\right)\nonumber\\
&-\alpha \mathrm{Tr}\left(\left(\rho_\mathrm{A}(t)\otimes\dot{\rho}_\mathrm{B}(t)\right)H_{\mathrm{AB}}\right).
\end{align}
Unlike the heat flux, the work rate on each system depends on the parameter $\alpha$.
This means that the work transferred between system A and B 
cannot be uniquely determined due to the ambiguity of $\alpha$.
However, the sum of these work rates is independent of the parameter $\alpha$ such as
\begin{align}
\dot{W}_\mathrm{A}+\dot{W}_\mathrm{B}
&=\mathrm{Tr}\left(\rho_\mathrm{A}(t)\dot{H}_\mathrm{A}(t)\right),
\label{def_work}
\end{align}
and so we can uniquely determine $\dot{W}_A+\dot{W}_B$.

\section{Two models of refrigerator}
\label{sec:model}
We discuss two refrigerator models to cool down low-energy qubits whose energy difference 
between ground and excited states
is much smaller 
than the thermal energy. 
One of them uses an extra qubit, and this approach requires many reset and pulse operations on the extra qubit.
This has already been realized experimentally
~\cite{liu2014protection,london2013detecting}.
The other is our proposal that does not require frequent pulses or reset, although we need two extra qubits.

When we have many low-energy qubits,
their dark states 
(which do not interact with the extra qubit) prevent them from being cooled down.
This problem was, however, discussed  and overcome in Ref.~\cite{PhysRevA.105.012613}. 
In order to make our models tractable, we thus consider only 
a single target qubit to be cooled down.

\subsection{Refrigerator with frequent reset: Refrigerator~I}
\label{subsec:Ham_prev} 
Let us consider the Hamiltonian of the target qubit and extra qubit. We assume that an energy of the former (latter) is much smaller (larger) than the thermal energy. The Hamiltonian is described as
\begin{align}
	H_{\mathrm D}(t) &= H_{1}+H_{2}(t)+H_{12}, \label{previoush} 
\end{align}
where $H_{1}$ is the Hamiltonian of the target qubit (Qubit~1), 
$H_{2}(t)$ is the Hamiltonian of the extra qubit (Qubit~2) 
with a driving field for a spin-lock 
operation~\cite{liu2014protection,london2013detecting}.  $H_{12}$ is the interaction Hamiltonian between them. 
They are assumed as 
\begin{align}
H_{1} &= \frac{\omega_{1}}{2}\sigma_{1z}, \\
H_{2}(t) &= \frac{\omega_2}{2}\sigma_{2z} 
          +\lambda \sigma_{2y}\cos(\omega_1 t),  \\
H_{12} &= g_1 \sigma_{1x}\sigma_{2z},\label{h12}
\end{align}
where $\omega_i$ 
is the resonant frequency (energy difference between the ground and excited states in frequency unit) of
Qubit~$i$,
and $\lambda$ is the strength of the driving field,
$g_{i}$ is the interaction strength 
between Qubit~$i$ and $(i+1)$.
$\sigma_{i\alpha}$ is a Pauli-$\alpha$ operator acting on Qubit~$i$, for example, $\sigma_{1x}=\sigma_x\otimes I$ and $\sigma_{2x}=I\otimes \sigma_x$. $I$ is the identity operator of dimension 2.

The GKSL master equation 
of the total system is given as \cite{breuer2002theory}
\begin{align}
    \frac{d\rho(t)}{dt}
&=-i\left[H_{\mathrm D},\rho(t)\right]+\mathcal{D}_2(\rho(t)), \label{old_master}
\end{align}
where 
\begin{align}
& \mathcal{D}_2(\rho(t)) \nonumber \\
&=\gamma_{2}\left(n(\omega_2)+1\right)\times
    \left(\sigma_{2-}\rho(t)\sigma_{2+}- \frac{1}{2}\left\{\sigma_{2+} \sigma_{2-},\rho(t)
\right\}\right)
\nonumber \\
&+ \gamma_{2} n(\omega_2)\times
   \left(\sigma_{2+}\rho(t)
   \sigma_{2-}-\frac{1}{2}\left\{\sigma_{2-}
   \sigma_{2+},\rho(t)\right\}\right).
\end{align}
$\gamma_2$ is an energy relaxation rate of the extra qubit,  
$n(\omega)=1/\left( 
e^{\left(\omega /T\right)}
-1\right)$ is the Bose-Einstein occupation number, $T$ is the temperature of the environment,
and
$\mathcal{D}_2(\rho(t))$ describes dynamics caused by the energy 
relaxation of Qubit~2. 
We assume that the energy relaxation of (low-energy) Qubit~1
is negligible.
For example, an electron spin has a long energy relaxation time, and so this assumption is reasonable for our purpose \cite{Budoyo_2018}.

By setting $\lambda = \omega_1$, the Hamiltonian $H_{\mathrm D}(t)$ provides a swapping interaction between Qubit~1 and 2, 
as shown in Appendix \ref{app:a}
\cite{hartmann1962nuclear,liu2014protection,london2013detecting}.
Thus, the energy can 
be transferred from Qubit~1 to Qubit~2. 
On the other hand,
when an energy relaxation rate of Qubit~2 is large,
Qubit~2 should be reset frequently.
The details of the protocol are explained in Appendix~\ref{app:b}.

\subsection{Refrigerator without frequent reset: Refrigerator~II}
\label{subsec:our}
We propose a quantum refrigerator without frequent reset 
by introducing the third qubit, Qubit~3. 
There is an interaction between Qubit~2 and 3. 
The energy of Qubit~3 is much larger than the thermal energy.
Moreover, unlike Qubit~2, we do not drive Qubit~3.
The Hamiltonian is now given as, 
\begin{align}
H_{\mathrm C}(t) 
&= H_1 + H_2(t)+H_3+ H_{12}+H_{23}, \\
H_3&= \frac{\omega_3}{2}\sigma_{3z}\\
H_{23} &= g_3\sigma_{2x}\sigma_{3x}, \label{h23}
\end{align}
where $H_3$ is the Hamiltonian of Qubit~3, $\omega_3$ is the resonant frequency of Qubit~3, and $g_3$ is the interaction strength between Qubit~2 and 3.
Here, $\sigma_{3z}=I \otimes I \otimes \sigma_z$,  $\sigma_{3x}=I \otimes I \otimes \sigma_x$, and $\sigma_{2x}=I\otimes \sigma_x \otimes I$.
The GKSL master equation for the total system is now given as,
\begin{align}
\frac{d\rho(t)}{dt}
&=-i\left[H_{\mathrm C}(t),\rho(t)\right]
+\mathcal{D}_2(\rho(t)) + \mathcal{D}_3(\rho(t)), \label{new_master}
\end{align}
where 
\begin{align}
&\mathcal{D}_3(\rho(t)) \nonumber \\
&=\gamma_3 \left(n(\omega_3)+1\right)\times 
\left(\sigma_{3-} \rho(t) \sigma_{3+}-\frac{1}{2}\left\{\sigma_{3+} \sigma_{3-},
\rho(t)\right\}\right) \nonumber \\
&+\gamma_3 n(\omega_3) \left(\sigma_{3+} \rho(t)\sigma_{3-}
-\frac{1}{2}\left\{\sigma_{3-}\sigma_{3+},\rho(t)\right\}\right) . 
\end{align}
$\gamma_3$ is the energy relaxation rate of Qubit~3, and  
$\mathcal{D}_3(\rho(t))$ describes dynamics caused by the energy 
relaxation of Qubit~3. 

We set
$\omega_3 = \omega_1 + \omega_2$ and $\gamma_2 \ll \gamma_3 \ll \omega_3$. 
The first condition is necessary in order to obtain an effective flip-flop 
interaction between Qubit~2 and 3.
Owing to this interaction, Qubit~3 can cool down Qubit~2 (see Appendix~\ref{app:a}).
The condition of $\gamma_2 \ll \gamma_3$ is required to keep Qubit~3 in a thermal
equilibrium state.
On the other hand, we need the condition of
$\gamma_3 \ll \omega_3$
so that
Qubit~3 should be a well-defined qubit although its energy relaxation 
is quite fast. In addition to these conditions, we need a condition of $\lambda = \omega_1$ 
to lead the flip-flop interaction between Qubit~1 and 2. In total, 
the energy of Qubit~1 is extracted and transferred to the environment 
through Qubit~2 and 3.

\section{Numerical Calculation}
\label{sec:result}
We show numerical calculations with realistic parameters for Qubit~1, 2 and 3. 

\subsection{Parameters}
We set $\omega_2$ as a scale and measure the other parameters by comparing with it. 
In Table~\ref{table:parameters}, 
the parameters, which are normalized by $\omega_2$, for the numerical calculations are summarized.
As shown in the Appendix \ref{app:a}, we consider  superconducting flux qubits to realize our proposal. 
Thus, $\omega_2$ is the order of GHz.

\begin{table}[htbp]
  \caption{The parameters normalized by $\omega_2$ for the numerical calculations.}
  \label{table:parameters}
  \centering
  \begin{tabular}{lcr}
    \hline
    $\omega_1$  & $1/10$ \\
    $\omega_3 =\omega_1+ \omega_2$& $11/10$ \\
    $\lambda=\omega_1$ & $1/10$ \\
    $g_1$  & $5/100000$  \\
    $g_3$  & $8/10000$ - $40/10000$ \\
    $\gamma_2$  &  $1/10000$  \\
    $\gamma_3$  &  $1/100$ \\
    $T$  &  $1/10$  \\
    \hline
  \end{tabular}
\end{table}

\subsection{Polarization}
\begin{figure}[htbp]
  \centering
  \includegraphics[width=8.5cm]{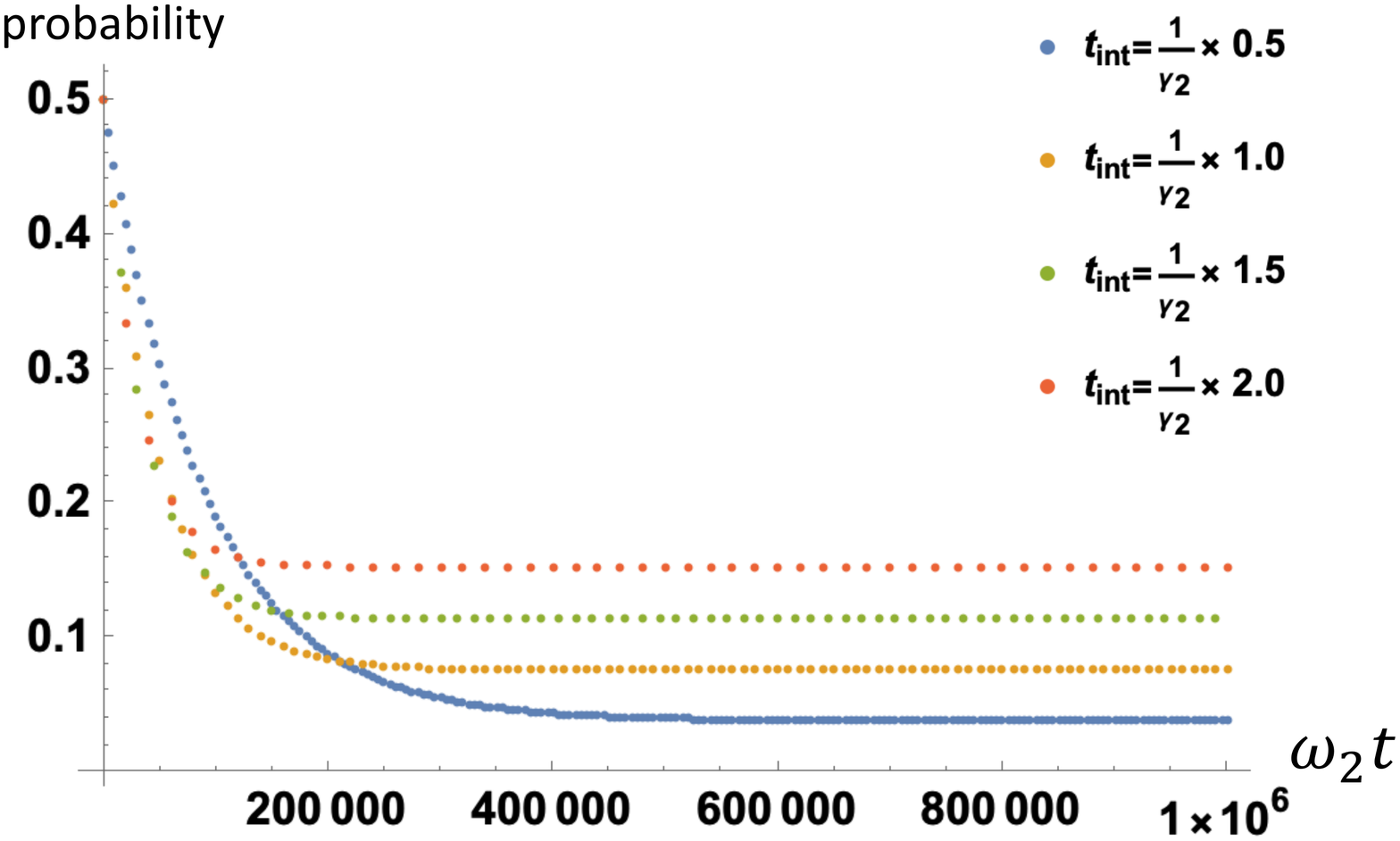}
  \vspace{-10mm}
  \caption{The probability of the excited state of Qubit~1 in 
  method of Refrigerator I. Qubit~1 is polarized due to the interaction with
  Qubit~2.}\label{fig_1}
\end{figure}

Here, we analyze the spin polarization in Refrigerator~I.
We calculate the population of the excited state $|1\rangle \langle 1|=\frac{I+\sigma_{1z}}{2}$ in Qubit~1 as follows,
\begin{align}
    \mathrm{Tr}\left(\rho^{(M)}(t_\mathrm{int})P_z\right)
\end{align}
where $\rho^{(M)}(t_\mathrm{int})$ is the total density matrix after the $M$-th reset, and $P_z=\frac{I+\sigma_{1z}}{2}$ is a projection operator to the excited state of Qubit~1.
In Fig.~\ref{fig_1}, we plot the probability of the excited state of Qubit~1.
In the calculation, we assume that the necessary time to initialize Qubit~2 is negligible compared to the interaction time $t_\mathrm{int}$.
As shown in Fig.~\ref{fig_1}, the probability of the excited state of Qubit~1 decreases in time.
Here, Qubit~1 is polarized due to the flip-flop interaction and the reset of Qubit~2 as explained in section \ref{sec:model}.
By fitting the population of the excited state in Qubit~1 with $a\left(e^{- t/T_\mathrm{cool}}-1\right)+1/2$, we determine the parameters $a$ and $T_\mathrm{cool}$, and estimate the cooling time $T_\mathrm{cool}$ for later calculations.

Next, we consider Refrigerator II where Qubit~3 is coupled to Qubit~2.
Similarly to Refrigerator I, we calculate the probability of the excited state in Qubit~1.
\begin{figure}[htbp]
  \centering
  \includegraphics[width=8.5cm]{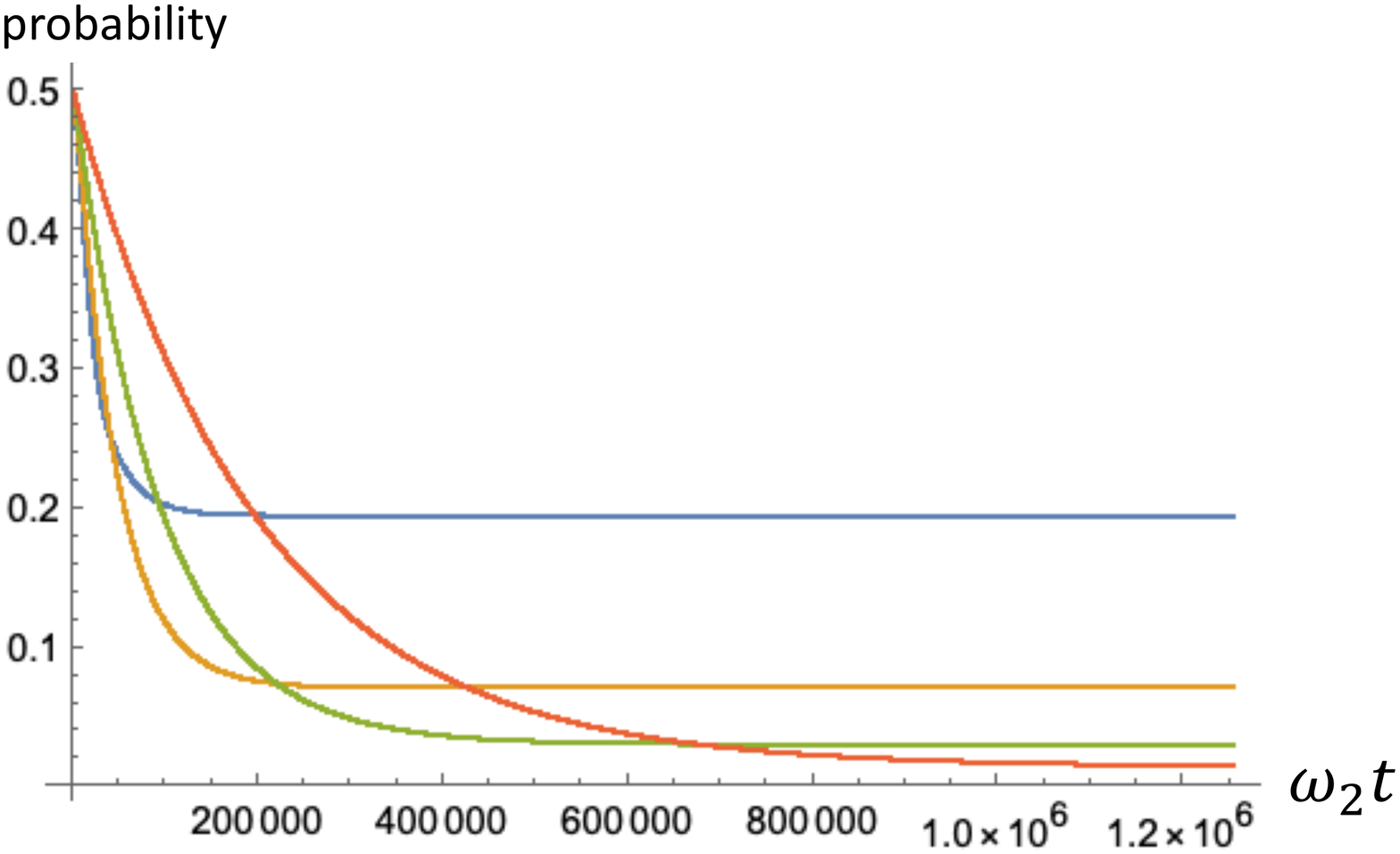}
  \vspace{-10mm}
  \caption{The probability of the excited state of Qubit~1 in the 
  method of Refrigerator II. The \{blue, orange, green, red\} line corresponds to the plot of $g_{3}/\omega_2=\{8.0,\ 15.0,\ 25.0,\ 40.0\}\times 1/10000$, respectively.}\label{fig_2}
\end{figure}
The heat from Qubit~1 is released to the environment via Qubit~3, and Qubit~1 is polarized continuously without the initialization of Qubit~2.
As the coupling constant $g_{3}$ becomes stronger, the probability of the excited state of Qubit~1 at the steady state becomes closer to zero.
As the coupling strength $g_3$ becomes smaller, the cooling process becomes faster.
Although the performance of Refrigerator II is comparable with that of Refrigerator I in the polarization rate and the cooling time, Refrigerator II polarizes Qubit~1 continuously without frequent reset of Qubit~2, which should be an advantage.

\subsection{Quantum thermodynamics}
We calculate the heat and the work for Refrigerator I and II according to \S~\ref{sec:def}.
In the case of Refrigerator~I, the total heat removed from Qubit~1, $Q_I^{\rm (out)}$, is evaluated as,
\begin{align}
Q_I^{\rm (out)} 
&=i\sum_M \int_0^{t_\mathrm{int}} dt 
\mathrm{Tr}\left(\left[H_{12},\chi^{(M)}(t)\right]H_1'^{(M)}(t)\right),
\label{heat:old}\\
    H'^{(M)}_1(t)
&=\frac{\omega_1}{2}\sigma_{1z}+\mathrm{Tr}_2\left(\left(I_1\otimes\rho_2^{(M)}(t)\right)H_{12}\right)\otimes I,
\end{align}
where $t_\mathrm{int}$ is a time interval between reset and 
superscript $^{(M)}$ indicates ``after the $M$-th reset''.
Note that $Q_I^{\rm (out)}$ has the opposite sign to the integration of the heat inflow, e.g., Eq. (\ref{def_heat}), because we are now interested in the heat {\it removed from} Qubit 1.
The total work done on the system is given as,
\begin{align}
W_I&=\sum_M\left(\int_0^{t_\mathrm{int}}dt \mathrm{Tr}\left(\rho_2^{(M)}(t)\dot{H}_2(t)\right)
+W_\mathrm{ini}\right)\label{work:old}\\
W_\mathrm{ini}&\sim  \frac{\omega_2}{2}. 
\end{align}
$W_\mathrm{ini}$ is the work needed to initialize Qubit~2 and is derived in Appendix \ref{app:c}.
Note that the total number of reset is given as $\lfloor 6T_\mathrm{cool}/\gamma_2 \rfloor$.

In the case of Refrigerator~II, the total heat removed from Qubit~1 is given as,
\begin{align}
Q_{II}^{\rm (out)}&=i \int_0^{\infty}dt 
\mathrm{Tr}\left(\left[H_{12},\chi(t)\right]H'_1(t)\right), 
\label{heat:new}\\
H'_1(t)
&=\frac{\omega_1}{2}\sigma_{1z}+\mathrm{Tr}_2\left(\left(I_1\otimes\rho_2(t)\right)H_{12}\right)\otimes I,
\end{align}
where $\chi(t)$ is the correlation between Qubit~1 and 2 and
$\rho_i(t)$ is the reduced density matrix of Qubit~$i$.
The total work done on the total system is given as
\begin{align}
W_{II} &=\int_0^{\infty}dt \mathrm{Tr}\left(\rho_2(t)\dot{H}_2(t)\right)+W_\mathrm{ini}.
\label{work:new}
\end{align}
As we will show later, $W_\mathrm{ini}$ is negligible compared to that 
during the implementation of the spin-lock.
In our numerical simulation, we integrate from 0 to $6 T_\mathrm{cool}$, because the system becomes almost a steady state at $t = 6 T_\mathrm{cool}$.

\begin{figure}[htbp]
  \centering
  \includegraphics[width=8.5cm]{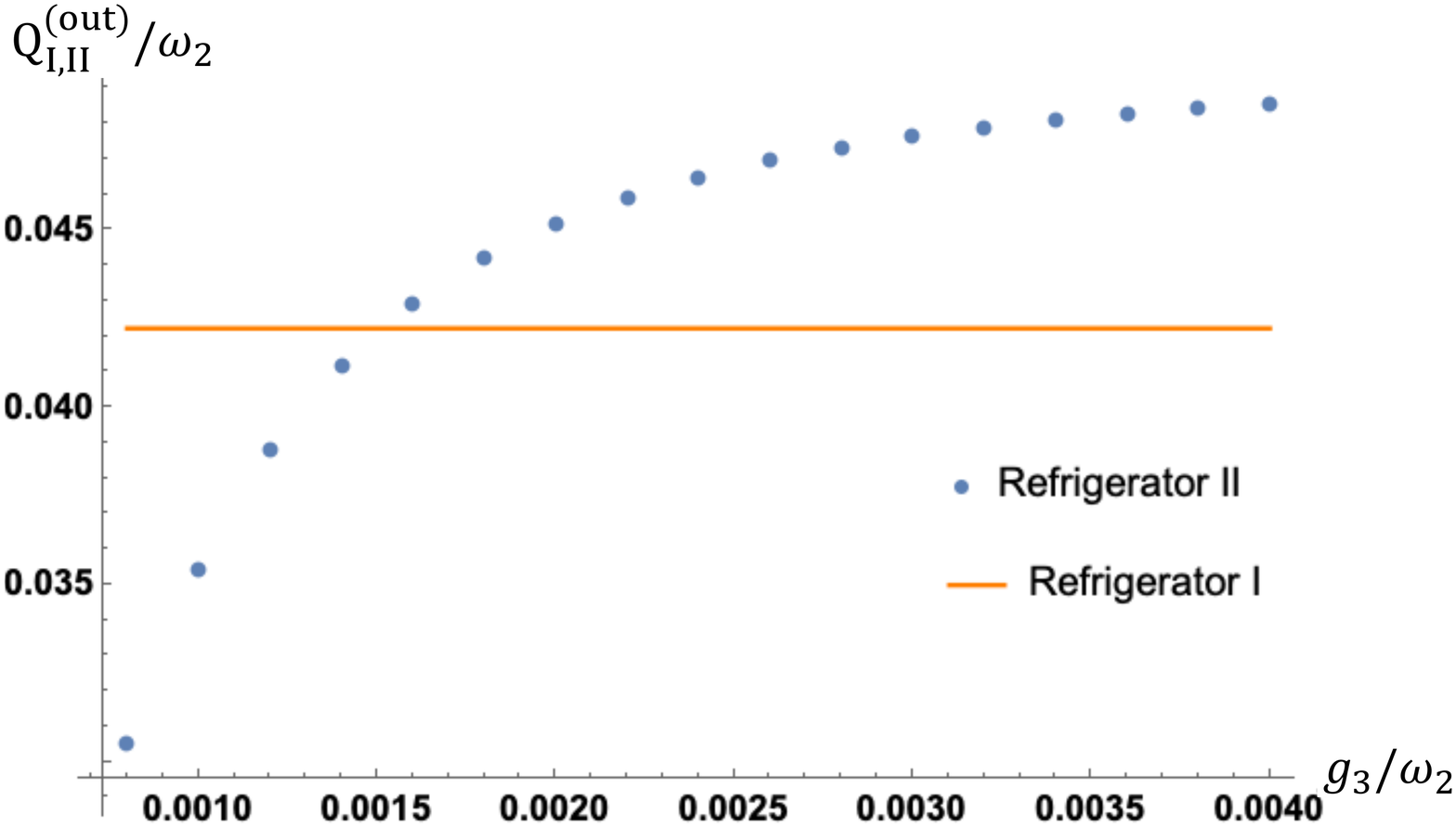}
  \vspace{-10mm}
  \caption{The heat removed from Qubit~1, $Q_{II}^{\rm (out)}$, 
as a function of the interaction strength between Qubit~2 and 3, $g_3$. 
The blue points show $Q_{II}^{\rm (out)}$,
while the orange horizontal line is $Q_{I}^{\rm (out)}$. }
\label{fig_3}
\end{figure}

In Fig.~\ref{fig_3} we plot $Q_{II}^{\rm (out)}$ as a function of $g_3$.
Note that, if the heat is extracted from Qubit~1, the sign of  $Q_{I,II}^{\rm (out)}$ is positive.
For $g_3/\omega_2> 0.0014$, $Q_{II}^{\rm (out)}$ of Refrigerator~II becomes larger than that of Refrigerator~I, $Q_{II}^{\rm (out)}>Q_{I}^{\rm (out)}$.
This means that, by tuning $g_3$, Refrigerator~II outperforms Refrigerator~I.

\begin{figure}[htbp]
  \centering
  \includegraphics[width=8.5cm]{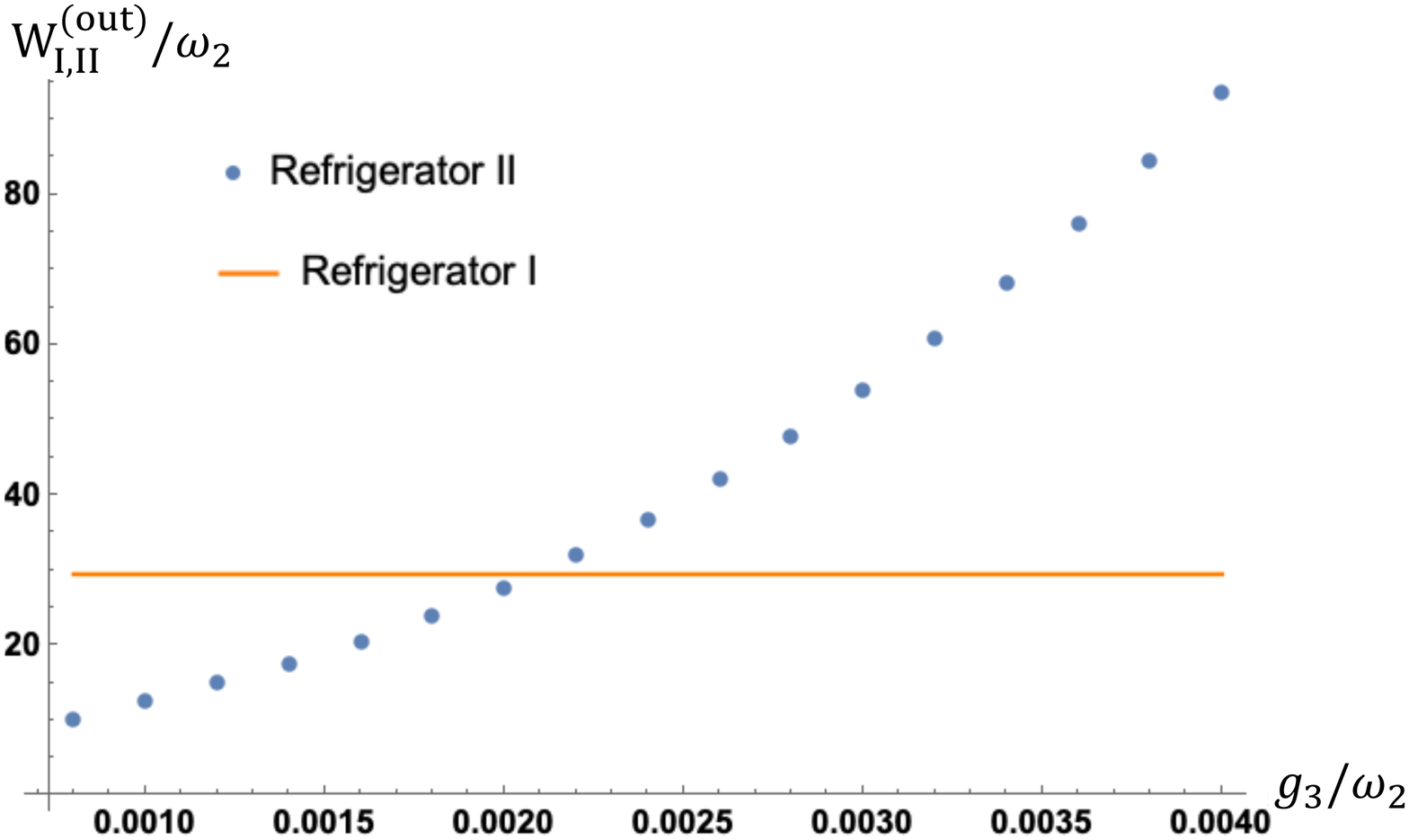}
  \vspace{-10mm}
  \caption{
The total work injected into the system, $W_{II}$, as a function of 
the interaction strength between Qubit~2 and 3, $g_3$. 
The blue points denote $W_{II}$, while the orange horizontal line denotes
$W_{I}$.}
\label{fig_4}
\end{figure}

Fig.~\ref{fig_4} shows the total work consumed for the
driving field and the initialization of Qubit~2 as a function of $g_3$.
The interaction time $t_\mathrm{int}$ is set to $1/\gamma_2$
in the case of Refrigerator~I.
Therefore, the total number of reset is $\lfloor 6T_\mathrm{cool}/\gamma_2 \rfloor$.
As a comparison, we also plot the total work for Refrigerator~I.
The total amount of the work in Refrigerator~I, $W_I/\omega_2$, is $29.42$, while the amount of the work for resetting Qubit~2 in Refrigerator~I, $\sum_M W_{\mathrm{ini}}/\omega_2$, is $14.50$.
For Refrigerator~I, almost half of the total work is devoted to initialize the qubit.
So the frequent reset is not desirable to realize an efficient
refrigerator which cools down the spin.

\begin{figure}[htbp]
  \centering
  \includegraphics[width=8.5cm]{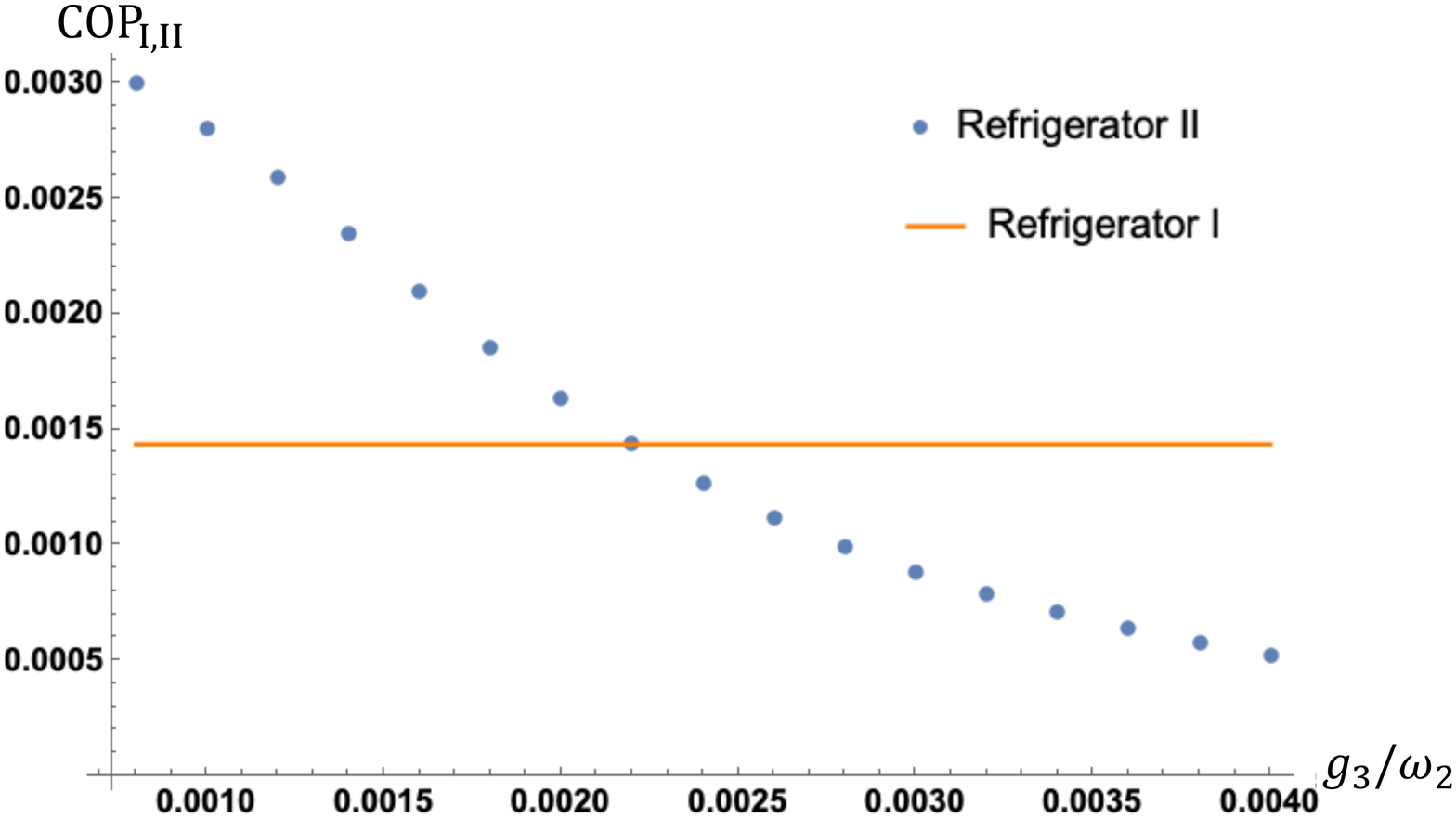}
  \vspace{-10mm}
  \caption{The coefficient of performance (COP) as a function of  the interaction strength 
between Qubit~2 and 3, $g_3$.  
The blue points denote $\mathrm{COP}_{II}$, while the orange horizontal line denotes 
$\mathrm{COP}_{I}$.}\label{fig_5}
\end{figure}

We evaluate the coefficient of performance (COP) defined as,
\begin{align}
    \mathrm{COP}_{I,II}=\frac{Q_{I,II}^{\rm (out)}}{W_{I,II}}.
\end{align}
The COP quantifies how large the heat is extracted from Qubit 1 by a unit of the invested work on Qubit 2,
and thus the COP should be larger for a better refrigerator.
Fig.~\ref{fig_5} shows the COP as a function of $g_3$, which is the interaction strength 
between Qubit~2 and 3.
Again, we show that, by tuning $g_3$, Refrigerator~II can be more efficient than the Refrigerator~I.

\begin{figure}[htbp]
  \centering
  \includegraphics[width=8.5cm]{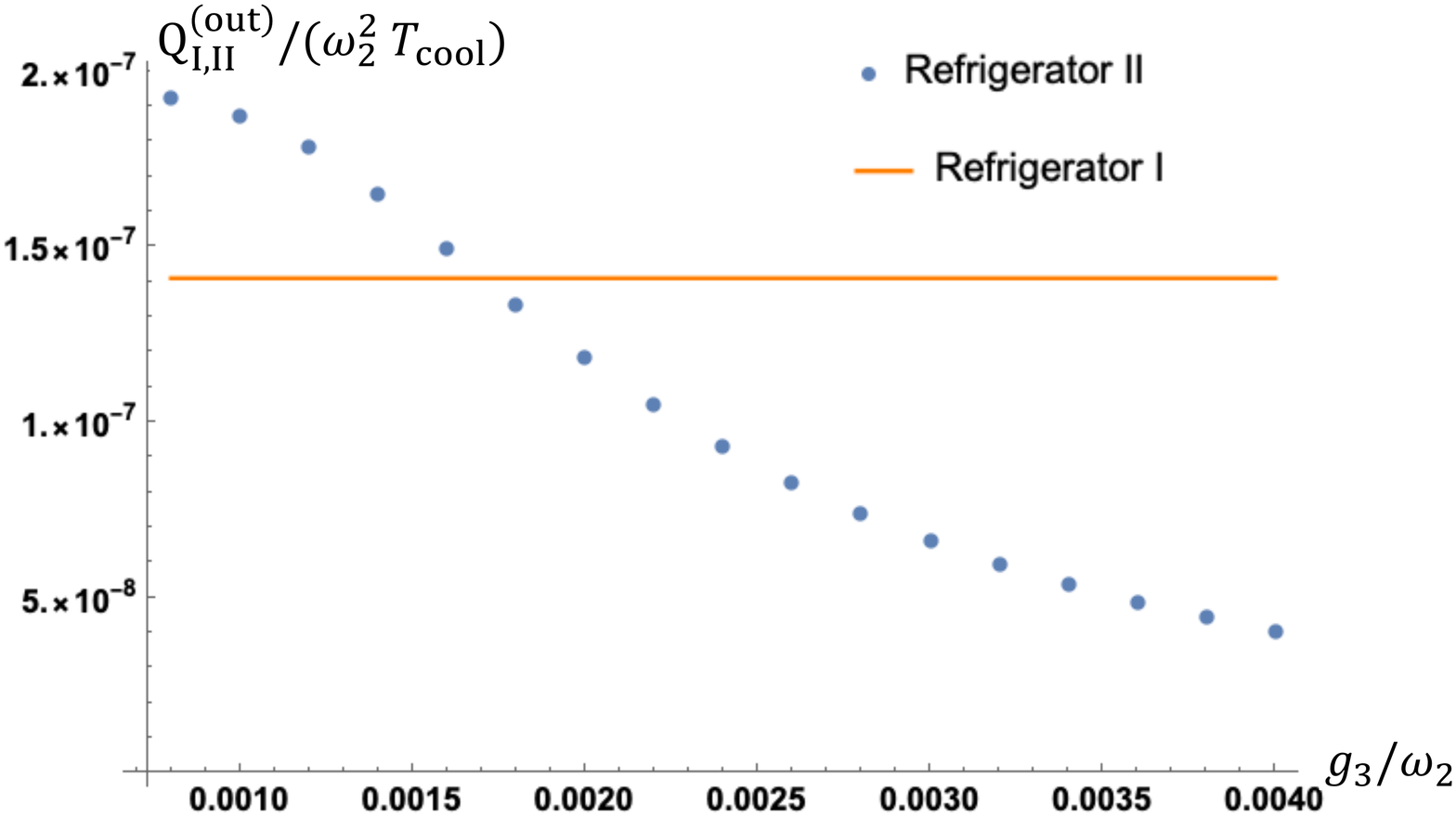}
  \vspace{-10mm}
  \caption{The average heat flow as a function of the interaction strength 
between Qubit~2 and 3, $g_3$.  
The blue points are the average heat flow of Refrigerator~II, while the orange 
horizontal line is that of Refrigerator~I. }
\label{fig_6}
\end{figure}

We evaluate the speed of the heat transfer and 
introduce the average heat flow as follows,
\begin{align}
\frac{Q_{I,II}^{\rm (out)}}{T_\mathrm{total}}
&=\frac{1}{T_\mathrm{total}}\int_0^{T_\mathrm{total}}dt'\ \dot{Q}_{I,II}^{(\mathrm{out})}(t')
\end{align}
where $T_\mathrm{total}=6T_\mathrm{cool}$.
Fig.~\ref{fig_6} shows the average heat flow  
as a function of 
$g_3$, which is the interaction strength between Qubit~2 and 3.
Although the polarization rate of the spin becomes better by increasing $g_3$ in Fig.~\ref{fig_3}, the speed becomes slower as we increase $g_3$ in Fig.~\ref{fig_6}.
This leads us to analyze the trade-off relationship between them.

\subsection{Trade-off relations}
\label{subsec:trade}
\begin{figure*}
  \begin{minipage}[b]{0.45\linewidth}
    \centering
    \includegraphics[width=8.5cm]{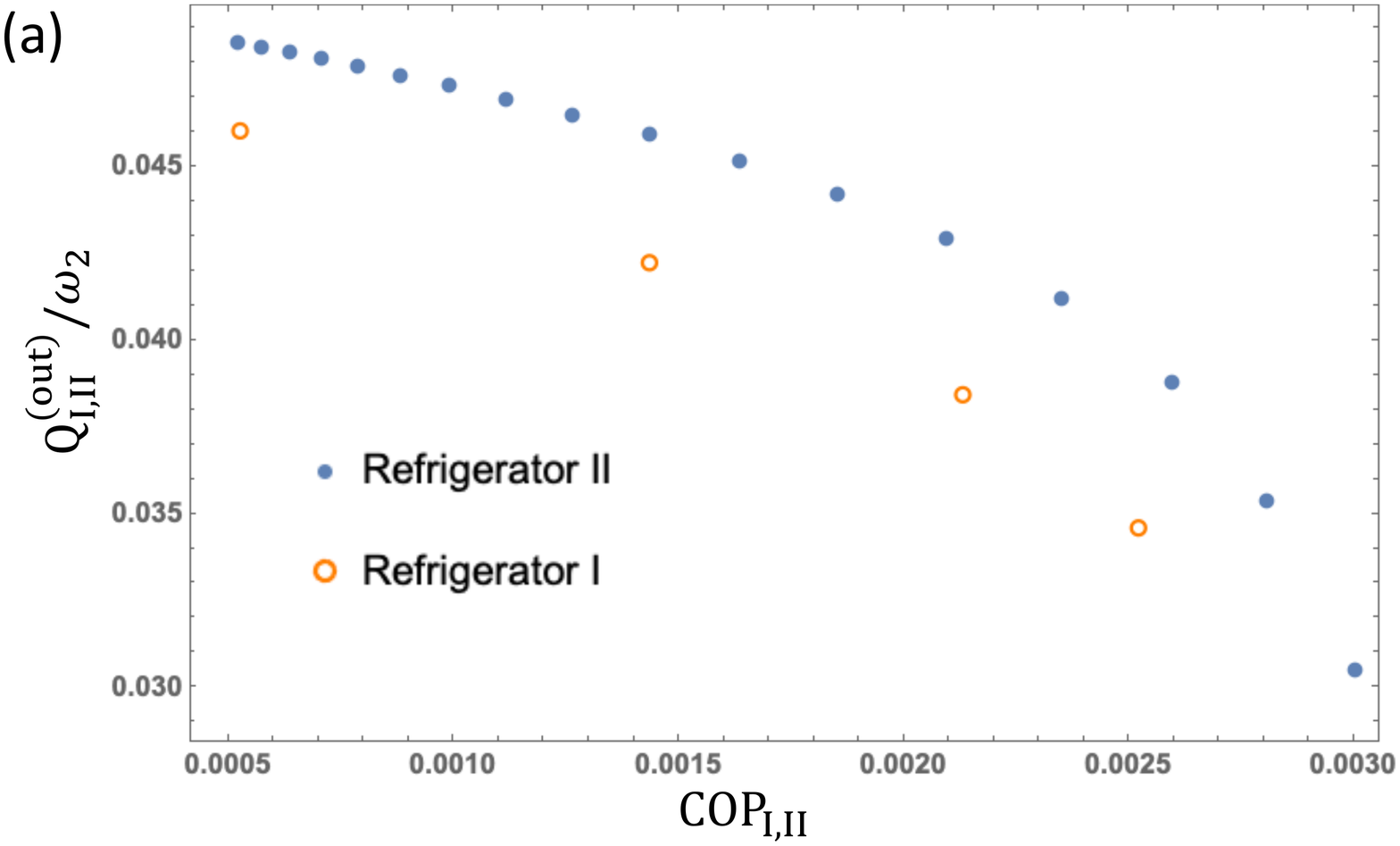}
  \end{minipage}
  \begin{minipage}[b]{0.45\linewidth}
    \centering
    \includegraphics[width=8.5cm]{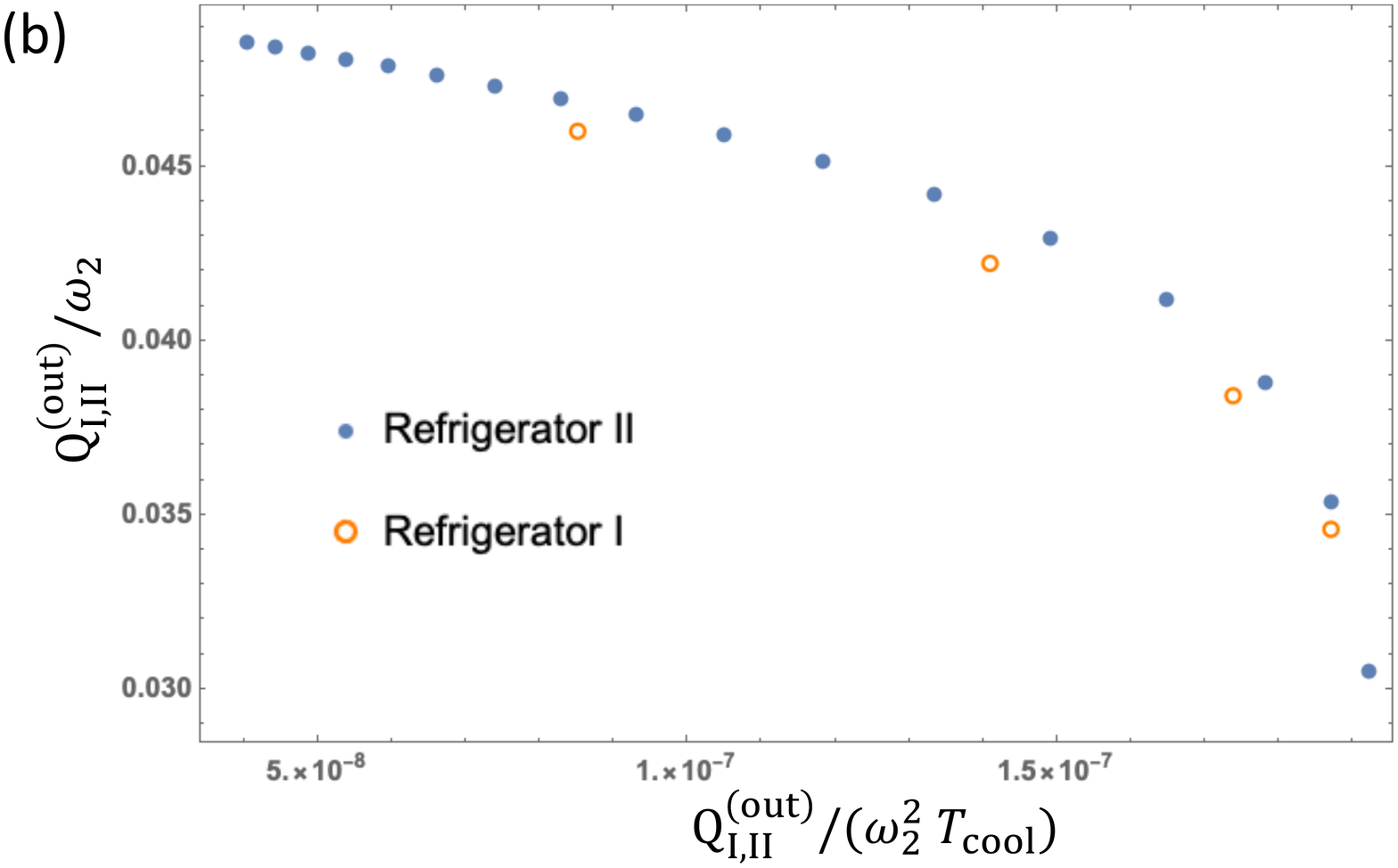}
  \end{minipage}\vspace{-10mm}
  \caption{The two trade-off relations about the heat removed from Qubit~1. 
The blue points are obtained by Refrigerator~II with the same interaction strength 
as in Fig.~\ref{fig_3}, Fig.~\ref{fig_5}, and Fig.~\ref{fig_6}. 
The orange open points are obtained by Refrigerator~I 
with $t_\mathrm{int}=\{1/(2\gamma_2),\ 1/\gamma_2,
\ 3/(2\gamma_2),\ 2/\gamma_2\}$. 
(a) The trade-off relation between the heat $Q^{\rm (out)}_{I,II}$ and $\mathrm{COP}_{I,II}$. 
(b) The trade-off between the heat $Q^{\rm (out)}_{I,II}$ and the average heat flow, 
$Q_{I,II}^{\rm (out)}/T_\mathrm{total}$.}
\label{fig_78}
\end{figure*}

We show two trade-off relations about the heat moved from Qubit~I.
The relation between the heat $Q_{I,II}^{\rm (out)}$ and $\mathrm{COP}_{I,II}$ is shown in Fig.~\ref{fig_78} (a),
while that between $Q_{I,II}^{\rm (out)}$ and the average heat flow, $Q_{I,II}^{\rm (out)}/T_\mathrm{total}$ 
in Fig.~\ref{fig_78} (b).
The former shows that, when we improve the COP, the total amount of the heat from Qubit~1 decreases.
On the other hand, the latter shows that, as we increase the total amount of the heat from Qubit~1, the cooling speed becomes worse.
This means that, by tuning $g_3$, we could cool down the spin with a high-polarization rate or we could quickly cool down the spin with a low-polarization rate.
We can choose a suitable strategy, depending on a purpose.
Although similar trade-off relationships were discussed for a cyclic discrete quantum refrigerator~\cite{PhysRevE.87.012105,PhysRevE.89.062119,PhysRevE.91.042127,KAUR2021125892,PhysRevE.106.024137,https://doi.org/10.48550/arxiv.2207.03374}, we show such a trade-off relationship about a continuous cooling scheme to polarize a spin.
These results show that,
due to the extra cost of the initialization of Qubit~2, the COP of Refrigerator~I becomes worse than that of Refrigerator~II, $\mathrm{COP}_{I}<\mathrm{COP}_{II}$.
This means that our proposed method is advantageous over the conventional one from the viewpoint of quantum thermodynamics.

\section{Summary}
\label{sec:summary}

In this work, we evaluate the performance of two types of quantum refrigerators cooling down a target qubit (Qubit~1) whose resonant frequency is much smaller than a thermal energy. One of them uses an extra qubit with frequent pulse operations, and the other uses two extra qubits without frequent pulse operations. 
The former has been experimentally realized 
\cite{liu2014protection,london2013detecting}.
The latter is a newly proposed method in this paper, where one of the extra qubits is spin-locked and the other is kept in an almost ground state because of a short $T_1$.
We evaluate these refrigerators 
from the viewpoint of quantum thermodynamics.
We find the two trade-off relations about the heat transferred from 
the low-energy qubit: the heat removed from the target qubit vs the COP, and the heat removed from the target qubit vs 
the average heat flow.
Furthermore,
we find that our proposed approach can be better than the one with frequent pulse operations in terms of 
reduction of the total work.
The advantage 
comes from the fact that our method requires less pulse operations.
Our results contribute to design high-performance quantum refrigerators which cool down a qubit.

\section*{Acknowledgment}
We thank helpful comments from Y. Mori and A. Yoshinaga.
This work was supported by Leading Initiative for
Excellent Young Researchers MEXT Japan, JST presto
(JPMJPR1919) Japan, JSPS Grants-in-Aid for Scientific Research (21K03423), 
and CREST (JPMJCR1774).

\appendix
\section{Realization of Refrigerator~II with superconducting flux Qubits }
\label{app:a}
Our proposal, which is inspired by experiments in Refs.~\cite{PhysRevB.102.100502,abdurakhimov2020driven},
can be realized by using two superconducting flux qubits (FQ1 and FQ2).
Here, Qubit~1 is an electron spin~\cite{PhysRevA.105.012613}.
Thus, $\omega_2$ is the order of GHz. In this work, we set $\omega_2/2\pi = 1.0$~GHz. 
In Table.~\ref{table:parameters}, 
the parameters, which are normalized by $\omega_2$ for the numerical calculations, are summarized.
The coupling strength between the flux qubit and an electron spin can be as large as tens of kHz
\cite{matsuzaki2015improving,twamley2010superconducting}.
The coupling strength between the flux qubits is an order of MHz
\cite{plantenberg2007demonstration}.
The driving strength of the flux qubit can be as large as hundreds of MHz \cite{yoshihara2014flux}.
The energy relaxation time of the flux qubits can be as short as tens of nano seconds
\cite{kakuyanagi2007dephasing}, while it can be an order of a few micro seconds \cite{bertet2005dephasing}, depending on the circuit design.  So all of these parameters are realistic even in the current technology.
For example, the parameters listed in Table~\ref{table:parameters} correspond to $\omega _1/2\pi= 0.1$ GHz, $\omega _2/2\pi= 1$ GHz, $\lambda/2\pi= 0.1$ GHz, $g _1/2\pi= 50$ kHz, $g _1/2\pi= 0.8 - 4$ MHz, $\gamma _2/2\pi = 0.1$ MHz, and $\gamma _3/2\pi = 10$ MHz.

We show that an energy from an electron (Qubit~1) can be transferred to the  FQ2 (Qubit~3) via the FQ1 (Qubit~2) by satisfying the resonant condition.
We will show the resonant condition below.
We assume that the interaction between the FQs and also that between the FQ and electron spin are inductive.
Therefore, the total Hamiltonian is as follows \cite{PhysRevLett.94.090501, doi:10.7566/JPSJ.91.064004}:
\begin{align}
H_1&= \frac{\omega_1}{2}\sigma_{1z},\\
H_2(t)
&= \frac{\omega_2'}{2}\sigma_{2z} +\frac{\Delta_2}{2} \sigma_{2x}
+\lambda \sigma_{2y}\cos{(\omega t)},\\
H_3&= \frac{\omega_3'
}{2}\sigma_{3z}+\frac{\Delta_3}{2}\sigma_{3x},\\
H_{12} &= g_1'\sigma_{1x}\sigma_{2z},\\
H_{23} &= g_3'\sigma_{2z}\sigma_{3z},
\end{align}
where $\omega_1$ denotes the Zeeman energy of the electron spin, $\omega_2'$ ($\omega'_3$) denotes an energy bias of the FQ1 (FQ2), $\Delta_2$ ($\Delta_3)$ the gap energy of the FQ1 (FQ2), $\lambda$ the rabi frequency, $g_1'$ the inductive coupling between the electron spin and FQ1, and $g_3'$ the inductive coupling between the FQ1 and FQ2.
We choose $\omega_3'=0$ by switching off the magnetic field applied to FQ2.
First, we diagonalize $\frac{\omega_2'}{2}\sigma_{2z} +\frac{\Delta_2}{2} \sigma_{2x}$ by 
$U=\exp{\left(i\theta \sigma_y/2\right)}$ with 
$\theta=\tan^{-1}\left(\Delta_2/\omega_2 \right)$.
We obtain 
\begin{align}
%
H_{\mathrm C}'(t)
&=H_1+ H_2'(t)+H_3+H_{12}'+H_{23}', 
\end{align}
where
\begin{align}
H_2'(t) 
&= \frac{\omega_2}{2}\sigma_{2z}+\lambda \sigma_{2y} \cos{(\omega_2 t)},
\\
H_{12}'
&=\frac{g_1'}{\omega_2}\sigma_{1x}\left(\omega_2'\sigma_{2z}-\Delta_2\sigma_{2x}\right),\label{h'12}
\\
H_{23}' &= 
\frac{g_3'}{\omega_2}\left(\omega_2' \sigma_{2z}-\Delta_2\sigma_{2x}\right) \sigma_{3z},\label{h'23}
\end{align}
where $\omega_2 =\omega = \sqrt{\omega_2'{}^2+\Delta_2^2}$.
Note that the interaction Hamiltonian of Eqs.~\eqref{h12} and \eqref{h23} correspond to the first term of Eq.~\eqref{h'12} and the second term of  Eq.~\eqref{h'23}, respectively.
By setting $g_1'\omega_2'/\omega_2=g_1$ and $-g_3'\Delta_2/\omega_2=g_3$, $\Delta_3=\omega_3$, and replacing $\sigma_{3z}$ by $\sigma_{3x}$, we obtain the correspondence to the model of the main text.
As seen below, the other terms
in ~\eqref{h'12} and ~\eqref{h'23} 
can be neglected by using the rotating wave approximation (RWA).

We move to the rotating frame defined by the unitary transformation 
$U_1(t)=\exp{\left(i\omega_2 t \sigma_{2z}/2\right)}
\exp{\left(i\omega_2 t \sigma_{3x}/2\right)}$.
Then, by using the RWA, we obtain
\begin{align}
H_{\mathrm C}''
&=\frac{\omega_1}{2}\sigma_{1z}+\lambda\left(\cos{\left(\omega_2 t\right)}\sigma_{2y}+\sin{\left(\omega_2 t\right)}\sigma_{2x}\right)\cos{(\omega_2 t)}\nonumber\\
&+\frac{\Delta_3-\omega_2}{2}\sigma_{3x}\nonumber\\
&+\frac{g_1'}{\omega_2}\sigma_{1x}\left(\omega_2'\sigma_{2z}-\Delta_2\left(\cos{\left(\omega_2 t\right)}\sigma_{2x}-\sin{\left(\omega_2 t\right)}\sigma_{2y}\right)\right)\nonumber\\
&+\frac{g_3'}{\omega_2}\left(\omega_2'\sigma_{2z}-\Delta_2\left(\cos{\left(\omega_2 t\right)}\sigma_{2x}-\sin{\left(\omega_2 t\right)}\sigma_{2y}\right)\right)\nonumber\\
&\times \left(\cos{\left(\omega_2 t\right)}\sigma_{3z}+\sin{\left(\omega_2 t\right)}\sigma_{3y}\right)\nonumber\\
&\simeq \frac{\omega_1}{2}\sigma_{1z}+\frac{\lambda}{2}\sigma_{2y} +\frac{\Delta_3-\omega_2}{2}\sigma_{3x}
\nonumber \\
&+\frac{g_1' \omega_2'}{\omega_2} \sigma_{1x}\sigma_{2z}-\frac{g_3' \Delta_2}{2 \omega_2} \left(\sigma_{2x} \sigma_{3z}-\sigma_{2y}\sigma_{3y} \right)
.
\end{align}
We impose the resonant conditions  $\Delta_3-\omega_2 = \omega_1 = \lambda$. 
Then, we move to another rotating frame defined by the unitary transformation 
$U_2(t)=\exp{\left(\frac{i\lambda t}{2} \sigma_{1z}\right)}
\exp{\left(\frac{i\lambda t}{2} \sigma_{2y}\right)} 
\exp{\left(\frac{i\lambda t}{2} \sigma_{3x}\right)} $
and we obtain,
\begin{align}
H_{\mathrm C}'''&=\frac{g_1' \omega_2}{\omega_2'} \left(\cos{(\lambda t)}\sigma_{1x}-\sin{(\lambda t)}\sigma_{1y}\right)\nonumber\\
&\times\left(\cos{(\lambda t)}\sigma_{2z}-\sin{(\lambda t)}\sigma_{2y}\right)
\nonumber\\
&-\frac{g_3' \Delta_2}{2 \omega_2'}\left(\cos{(\lambda t)}\sigma_{2x}+\sin{(\lambda t)}\sigma_{2z}\right)\nonumber\\
&\times\left(\cos{(\lambda t)}\sigma_{3z}+\sin{(\lambda t)}\sigma_{3y}\right)
\nonumber\\
&+\frac{g_3' \Delta_2}{2 \omega_2'}\sigma_{2y}\left(\cos{(\lambda t)}\sigma_{3y}-\sin{(\lambda t)}\sigma_{3z}\right)\nonumber\\
&\simeq \frac{g_1' \omega_2}{2\omega_2'}\left(\sigma_{1x}\sigma_{2z}+\sigma_{1y}\sigma_{2x} \right)
 -\frac{g_3'\Delta_2}{4\omega_2'} \left(\sigma_{2x}\sigma_{3x}+\sigma_{2z}\sigma_{3y}\right)
\nonumber \\
&=
\frac{g_1'\omega_2}{\omega_2'} \left(\sigma_{1-}\sigma_{2y+} + \sigma_{1+}\sigma_{2y-}\right)\nonumber\\
&+\frac{i g_3' \Delta_2}{2\omega_2'} \left(\sigma_{2y+}\sigma_{3x-}-\sigma_{2y-}\sigma_{3x+}\right)
\label{eq:swap_H}
\end{align}
where $\sigma_{y\pm}=\ket{\pm _y}\bra{\mp_y}$,  $\ket{\pm_y}=\frac{\ket{0}\pm i\ket{1}}{\sqrt{2}}$, $\sigma_{x\pm}=\ket{\pm _x}\bra{\mp_x}$, and $\ket{\pm_x}=\frac{\ket{0}\pm \ket{1}}{\sqrt{2}}$.
Eq.~\eqref{eq:swap_H} implies that energy can be removed from the electron spins to the FQ~2 via the FQ~1.

\section{Protocol of Refrigerator~I}\label{app:b}
Here, we explain a protocol of Refrigerator~I.
We assume that Qubit~1 is prepared as a completely mixed state while Qubit~2 is prepared as $\ket{-_y}$. The goal is to generate a ground state of $| 1 \rangle$ for Qubit~1.
The method consists of the following steps:
\begin{enumerate}
\item 
Let the total system evolve by the GKSL master equation for a time $t_\mathrm{int}$ with a given initial state $\ket{-_y}$ of Qubit~2.
\item Qubit~2 is reset. 
More specifically, the state of the 
Qubit~2 is
projected into
 $|1\rangle$ or $|0\rangle$ by a measurement, and is rotated to $\ket{-_y}$ by performing a $\pm \pi/2$ pulse where the sign depends on the measurement results.
\item Repeat the step 1 and 2, $M$ times.
\end{enumerate}
A completely mixed state of Qubit~1 can be polarized after these steps.
More specifically, we write the total density matrix after the $i$-th reset as $\rho^{(i)}(t)$.
As the first step, the initial state of the total system is prepared as
$\rho^{(0)}(0)=\frac{I}{2}\otimes \ket{-_y}\bra{-_y}$.
Then, the spin-locking drive is applied for Qubit~2 during the interaction time $t_{\mathrm{int}}$ and the total state becomes $\rho^{(0)}(t_{\mathrm{int}})$.
In this step, Qubit~1 will be partially polarized.
In the second step, Qubit~2 is prepared in $\ket{1}$ or $\ket{0}$.
This can be done by a projective measurement.
Before the third step, we apply the $\frac{\pi}{2}$ pulse to Qubit~2 to obtain an initial state for the subsequent cycle $\rho^{(1)}(0)= \mathrm{Tr}_2\left[\rho^{(0)}(t_{\mathrm{int}})\right]\otimes \ket{-_y}\bra{-_y}$ (see Appendix \ref{app:c} for details).
We repeat these cycles $M$ times so that we can get the polarized spin state $\mathrm{Tr}_2\left[\rho^{(M)}(t_{\mathrm{int}})\right]$.

\section{The work for initializing Qubit~2}
\label{app:c}
We calculate the work $W_\mathrm{ini}$ necessary for initializing Qubit~2.
The procedure of the initialization consists of two steps:
(1) the projection measurement with $\sigma_{2z}$, 
(2) the conditional operation $\pm \pi /2$ rotation along the $x$-axis
according to the projective measurement result of $\pm 1$. 
Then, we obtain $\ket{-_y}\bra{-_y}$.

We assume that a required work for the projective measurement 
is negligible compared to the other work.
It is not straightforward to quantify the required work for the projective measurements.
It strongly depends on how we experimentally realize the projective measurements.
Anyway if we include the work of projective measurements, the performance of Refrigerator~I becomes worse, while the performance of Refrigerator~II is almost the same.
Even in such a case the COP of Refrigerator~II is still better than that of Refrigerator~II, and so our results to show the superiority of the Refrigerator~II over Refrigerator~I are unchanged.

Then, the work of the $\frac{\pi}{2}$ rotation is regarded as $W_\mathrm{ini}$.
The time required for the initialization is assumed to be much shorter 
than any other dynamics, such as relaxation, and thus we only consider 
$\frac{\omega_2}{2} \sigma_{2z} - \lambda' \cos (\omega_2 t) \sigma_{2x}$ term in the Hamiltonian.
Moreover, it is reduced to $-\frac{\lambda'}{2}\sigma_{2x}$ 
in the rotating frame whose frequency is $\omega_2$. 

If the initial state is assumed as $\rho(0)=\ket{1}\bra{1}$, 
the dynamics of Qubit~2 is given as 
\begin{align}
\rho'(t)
&= e^{i\frac{\lambda' t}{2}\sigma_{2x}} \ket{1}\bra{1} 
e^{-i\frac{\lambda' t}{2}\sigma_{2x}}
\nonumber \\
&= \frac{1-\cos(\lambda' t)\sigma_{2z}-\sin(\lambda' t) \sigma_{2y}}{2}.
\end{align}
From the above, the required initialization time is $\frac{\pi}{2\lambda'}$.
We, now, calculate $W_\mathrm{ini}$ with the work rate defined by Eq.~\eqref{def_work}. 
$W_\mathrm{ini}$ is given as, 
\begin{align}
W_\mathrm{ini}
&\simeq 
\int_0^{\frac{\pi}{2\lambda'}} dt' \mathrm{Tr}\left(\rho'(t')
\left( -\frac{\lambda' \omega_2}{2} \sigma_{2x}\right)\right)
=\frac{\omega_2}{2}
\label{Wini}. 
\end{align}
under the rotating wave approximation.


\bibliography{ref} 
\bibliographystyle{apsrev4-2}

\end{document}